# Band magnetism with inter-site correlations and interactions

J. Mizia, G. Górski and K. Kucab

Faculty of Mathematics and Natural Sciences,

University of Rzeszów, 35-959 Rzeszów, Poland

**Abstract**

We introduce the Hamiltonian to describe narrow band electrons. The physics of driving forces towards ferromagnetism is re-examined. Using different approximations it has been shown that the magnetic moments created by inter-site interaction and inter-site kinetic correlation decrease quickly with temperature. As a result of these interactions and the realistic density of states (DOS) the Curie temperatures calculated after fitting magnetic moments to their low temperature values are realistic. In the past the Curie temperatures calculated using only the on-site interaction were much higher than the experimental temperatures.

## 1. Introduction

The basic problem of itinerant magnetism of 3d transition elements is re-examined. The simplest way of introducing metallic ferromagnetism is the Stoner model which assumes a shift of the spin bands (proportional to magnetization) without changing their shape. This shift is caused by the Stoner exchange interaction between d electrons, $F_{in}$. The critical value of this interaction is obtained from the well known Stoner criterion for ferromagnetism:

$F_{in} > 1/\rho(\varepsilon_F)$. The same result is generated in the Hartree-Fock approximation (H-F) applied to the classic Hubbard model with the on-site interaction $U$; $U_{cr} > 1/\rho(\varepsilon_F)$.

The defect of such an approach is that magnetic moment created by the Stoner exchange interaction is "hard" with temperature. As a result, the calculated Curie temperature is much higher than the experimental value [1]. This negative effect is especially strong in the case of „flat" density of states (DOS).

Based on the previous papers [2-4], and on our two-pole approximation [5] we propose the model in which the width and shape of the spin bands changes with the occupationally depending hopping amplitude. In this model the ferromagnetism is supported by the decrease in kinetic energy rather than by reduction of potential energy. The model Hamiltonian is analyzed by the equation of motion Green Function method. The inter-site kinetic correlations depending on hopping interactions are included in the decoupling process together with the on-site correlations responsible for the Coulomb repulsion. The shapes of spin bands are modified and the width of majority spin band is reduced with respect to the minority spin band. The authors show that including the inter-site kinetic correlation and the inter-site interactions into the physics of magnetism allows us to obtain the magnetic moments at zero temperature and at the same time the realistic Curie temperatures. The on-site interactions, which in the H-F approximation do not contribute to these changes, are rather responsible for the local moments which do not disappear at the Curie temperature. On the other hand, their ordering is governed by a much weaker inter-site interaction, analyzed in this paper, giving realistic Curie temperatures.

## 2. Model Hamiltonian

The single band model Hamiltonian for itinerant electrons has the following form [2,3]:

$$H = -\sum_{<ij>\sigma} t_{ij}^{\sigma} c_{i\sigma}^{+} c_{j\sigma} + \frac{U}{2}\sum_{i\sigma} \hat{n}_{i\sigma}\hat{n}_{i-\sigma} - F_{in}\sum_{i\sigma} n_{\sigma}\hat{n}_{i\sigma}, \qquad (1)$$

where the operator $c_{i\sigma}^{+}(c_{i\sigma})$ is creating (annihilating) an electron with spin $\sigma = \uparrow, \downarrow$ on the $i$-th lattice site; $\hat{n}_{i\sigma} = c_{i\sigma}^{+} c_{i\sigma}$ is the electron number operator for electrons with spin $\sigma$ on the $i$-th lattice site. The interactions are: $U$ - on-site Coulomb repulsion, $F_{in}$ - single site Hund's type exchange interaction (interaction between different orbitals on the same lattice site in the multi-orbital single band model). The spin dependent correlated hopping integral between the $i$-th and $j$-th lattice site, $t_{ij}^{\sigma}$, is expressed by the occupation of sites $i$ and $j$ in the operator form

$$t_{ij}^{\sigma} = t_{ij}^{0}(1-\hat{n}_{i-\sigma})(1-\hat{n}_{j-\sigma}) + t_{ij}^{1}[\hat{n}_{i-\sigma}(1-\hat{n}_{j-\sigma}) + \hat{n}_{j-\sigma}(1-\hat{n}_{i-\sigma})] + t_{ij}^{2}\hat{n}_{i-\sigma}\hat{n}_{j-\sigma} \quad , \tag{2}$$

where $t_{ij}^{0}$ is the hopping amplitude for an electron of spin $\sigma$ when there are no other electrons in the Wannier states at sites $i$ and $j$. Parameter $t_{ij}^{1}$ is the hopping amplitude for an electron of spin $\sigma$ when one of the sites $i$ or $j$ is occupied by an electron with opposite spin. Parameter $t_{ij}^{2}$ is the hopping amplitude for an electron of spin $\sigma$ when both sites $i$ and $j$ are occupied by electrons with opposite spin.

To reduce the number of free parameters we assume additionally that

$$\frac{t_{ij}^{1}}{t_{ij}^{0}} = \frac{t_{ij}^{2}}{t_{ij}^{1}} = S < 1 \quad . \tag{3}$$

As a result the inter-site interactions are represented by the single parameter $S$.

### 3. Models of ferromagnetism

To obtain changes in the shape of density of states (DOS) through electron correlations, which already appear in the H-F approximation, we include the nonzero inter-site kinetic correlation, $I_{\sigma} = \langle c_{i\sigma}^{+} c_{j\sigma} \rangle$, and nonzero inter-site interactions represented by parameter $S$. This approach adds the shape changes of spin bands to the Stoner exchange shift. The majority spin band is narrowed and increased around the Fermi level which allows it to accommodate more electrons. This effect supports magnetism and is confirmed by the experimental evidence [2].

In the present paper we calculate magnetization dependence on temperature (see Fig. 1). Calculations are performed with the inter-site interactions ($S = 0.6$) and without them ($S = 1$). We use the H-F approximation and a recently developed high level two-pole approximation [5]. This new two pole approach is used simultaneously for the Coulomb interaction and the hopping interactions. It allows investigation of the influence of the inter-site interactions on ferromagnetism. The parameters used in the present calculations represent cobalt in the single band model, i.e. $n = 1.65$, $m = 0.344$ and half-bandwidth $D = 2.65$eV [1]. The Coulomb interaction $U = 3D$ is treated in the two-pole approximation [5]. Effects of multi orbital interactions are represented by the single site Hund's exchange interaction $F_{in}$ fitted (also in this approximation) to the spontaneous magnetization $m = 0.344$ at zero temperature. We obtained $F_{in} = 1.23$eV for $S = 0.6$ and $F_{in} = 1.53$eV for $S = 1$. It is smaller than the half band width. After taking into consideration five fold degeneracy of the $d$ band: $F_{in} = 5F_{in,0}$, we find that the exchange interaction for one orbital is $F_{in,0} \sim 0.2 - 0.3 eV$. This estimate is comparable with previous evaluations (see e.g. [6]). We have used the asymmetric DOS with parameter $a_1 = -0.7$ resembling the fcc DOS (see [7]). Due to the peak in the DOS around Fermi energy such a DOS favors magnetic alignment (see eg. [7,8]) and decreases values of Curie temperature even in the H-F approximation. The Curie temperature is reduced to $T_C = 2470$K without the inter-site correlations (dot-dashed green line in Fig. 1) and to $T_C = 1388$K with the inter-site correlations (dashed red line in Fig. 1).

At the symmetric DOS and still in the H-F approximation we obtain for cobalt, Curie temperature with inter-site correlations ($S = 0.6$) value 3300K (instead of 1388K) and without inter-site correlations ($S = 1$) a value 4710K (instead of 2470K), see [1].

Using the higher level two-pole approximation [5] reduces $T_C$ even further. The black solid curve shows the case with inter-site interactions, asymmetric DOS, high level approximation; the $T_C$ is reduced from 1388K to 520K. The blue dotted curve is for the case without inter-site interactions, asymmetric DOS, high level approximation; the $T_C$ is reduced from 2470K to 1840K.

We can see that the inter-site correlations combined with the asymmetric DOS (dashed red line and green dot-dashed line in Fig. 1) reduce $T_C$ strongly allowing us to obtain critical temperatures close to the experimental value. These temperatures are reduced even further when the higher level two-pole approximation is used. The 'magnetic paradox' characteristic for the flat DOS and the single site interactions disappears.

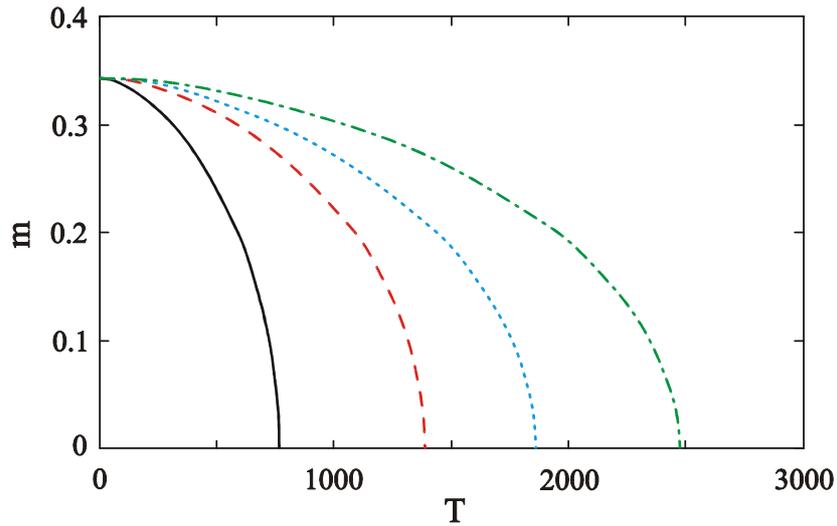

Fig. 1 Magnetization dependence on temperature, $m(T)$, without and with strong Coulomb correlation, $n=1.65$, $D = 2.65$eV for different values of parameter $S$, and Coulomb repulsion $U$; $U = 3D$, $S = 0.6$ - solid (black) line, $U = 0$, $S = 0.6$ - dashed (red) line, $U = 3D$, $S = 1$ - dotted (blue) line, $U = 0$, $S = 1$ - dot-dashed (green) line.

## 4. Conclusions

The inter-site kinetic correlations enhance ferromagnetism created by the inter-site interactions. This ferromagnetism is soft with temperature. Also the shape of the DOS peaked around the Fermi level helps in creating magnetic moments at zero temperature with a smaller exchange field. These two effects together remove the so called 'ferromagnetic paradox' of the Curie temperature being much too high after fitting the strength of on-site interaction to the zero temperature magnetic moments.

Our conclusions agree with the Hubbard's [9] calculations which 'imply that two energy scales are operative in iron, one of the order of electron volts which is characteristic of the itinerant behavior (e.g., the bandwidth and the exchange fields), and another of the order of one tenth of an electron volt characteristic of the "localized" behavior (e.g., $k_B T_C$, the $\Delta E(V)$)'. In today's model this larger field would be coming from on-site interactions which create local moments existing even above the Curie temperature, and the smaller field would be the inter-site field coming from the inter-site interactions and responsible for moments ordering.